\input{aipcheck}

\documentclass[
    ,final            
  ]
  {aipproc}
\layoutstyle{8x11single}
\usepackage{todonotes}
\usepackage{subfig}
\usepackage{caption}
\usepackage{float}
\usepackage{csquotes}
\MakeOuterQuote{"}
\usepackage{amsmath}

\begin{document}

\title{Geant4 simulations of radio signals from particle showers for the SLAC T-510 experiment }


\author{A.~Zilles\footnote{Email address: anne.zilles@kit.edu} { }for the SLAC T-510 Collaboration}{
address={Karlsruhe Institute of Technology (KIT), Germany }
  }

\begin{abstract}
The SLAC T-510 experiment \cite{Katie}  was designed to reproduce the physics of radio emission from air showers caused by ultra-high energy cosmic rays in a controlled lab experiment with the goal to test established formalisms for simulation of radio emission physics: 
the \enquote{end-point} formalism and the \enquote{ZHS} formalism. Simulation results derived with these formalisms can be explained by a superposition of magnetically induced transverse current radiation and the Askaryan (charge-excess) effect. 
Here, we present results of Geant4 simulations of the experiment with both formalisms, taking into account the details of the experimental setup (beam energy, target geometry and material, magnetic field configuration, and 
refraction effects) to test this hypothesis.
\end{abstract}

\maketitle

\section{Motivation}
\noindent
To understand actual models of air showers caused by ultra-high energy cosmic rays, we need to compare measured data with results from detailed Monte-Carlo air shower simulations.
The two main simulation tools available for  extensive air showers are AIRES (AIR-shower Extended Simulations) \cite{AIRES} and CORSIKA (COsmic Ray SImulations for KAscade) \cite{CORSIKA}.
These simulation tools require the continuous trajectories of the particles as they split in to multiple subtracks.
The positions of the end-points of these subtracks and their corresponding times are then available by design and can be used as the basis for the calculation of radio emission from particle showers. 
These Monte Carlo simulations make no assumptions on the mechanism of radio emission.
Two formalisms to calculate the electric field of the radio emission based on the movement of charged particles
are the \enquote{end-point} formalism \cite{Endpoint}\cite{REAS3} applied in CoREAS \cite{CoREAS} based on CORSIKA and the \enquote{ZHS} formalism \cite{ZHS} applied in ZHAireS \cite{ZHAireS} based on AIRES. 
An experiment under controlled laboratory conditions is needed to validate these models and to test for measurable differences between them suggested by previous discussions \cite{Alvarez2014}.\\
\\
\noindent
The T-510 experiment at the SLAC National Accelerator Laboratory aims to verify the radio emission mechanism by particle showers and to provide a benchmark to compare the predictions by the Monte Carlo simulations using these formalisms. 
The overview and the experimental setup are described in \cite{Katie}. 
The SLAC electron beam delivers bunches of electrons that produce showers which develope in a dielectric target and mimic those produced by individual higher energy primary particles of known species and energy.
The development of the induced shower takes place in a strong controllable magnetic field to study the variation of the radio signal produced as a function of magnetic field strength and polarity. 
Radio signals produced in the target are recorded with antennas positioned on a vertical axis with respect to the beam axis such that the measured horizontal component of the electrical field can be interpreted as the contribution of the 
\enquote{geo}-magnetic effect \cite{Geomagnetic}, i.e. a linearly polarised electrical 
field proportional to the cross product of the moving direction of the shower and the magnetic field vector, and the vertical one as the contribution of the Askaryan effect \cite{Askaryan},
i.e. a linearly polarised electric field which is radially oriented around the shower axis.\\
\section{Simulation scheme and results}
\begin{figure}[t]

\vbox{
\centering{
\includegraphics[width=0.32\textwidth]{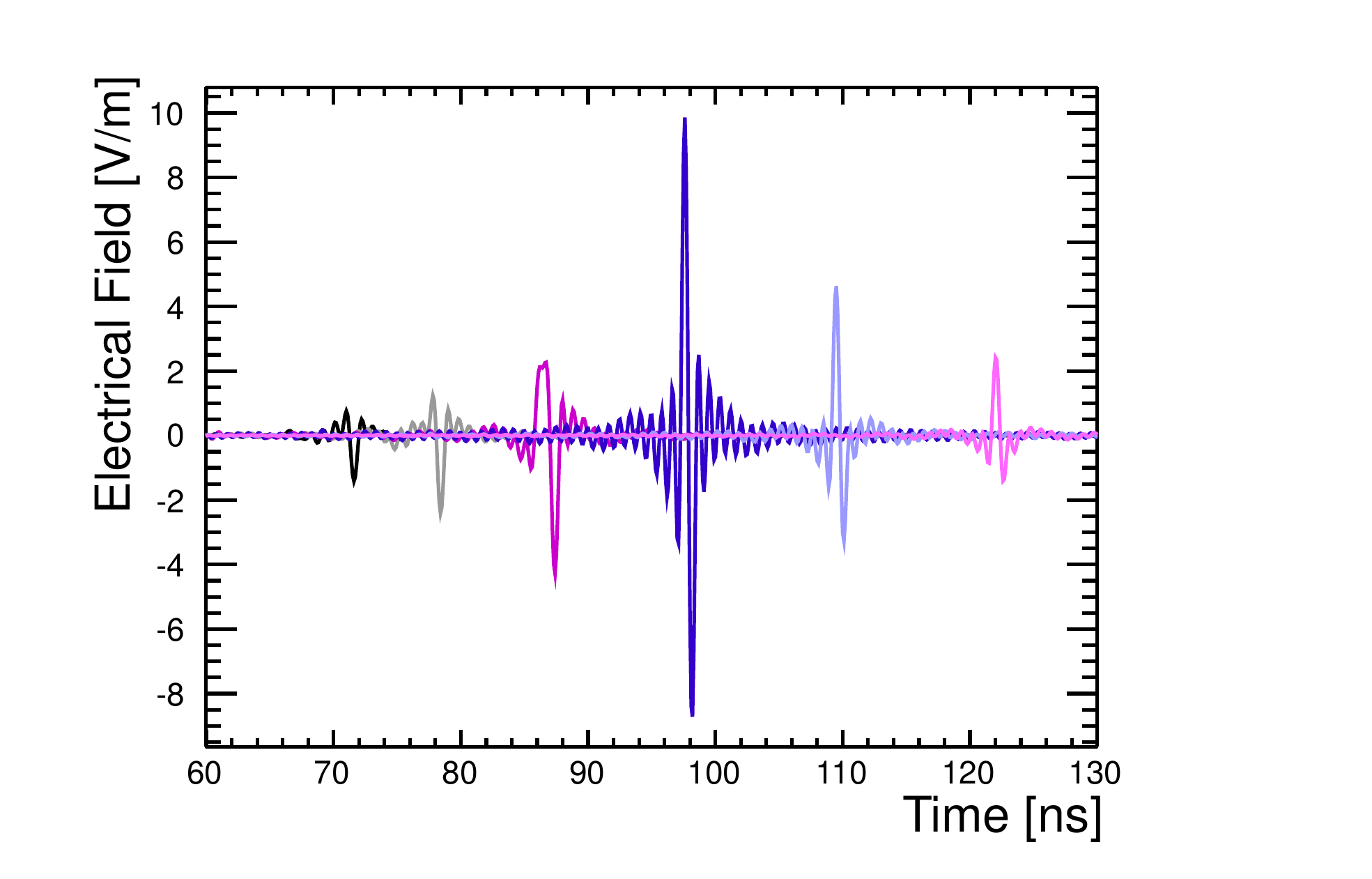}
%
\hspace{0.cm}
%
\includegraphics[width=0.32\textwidth]{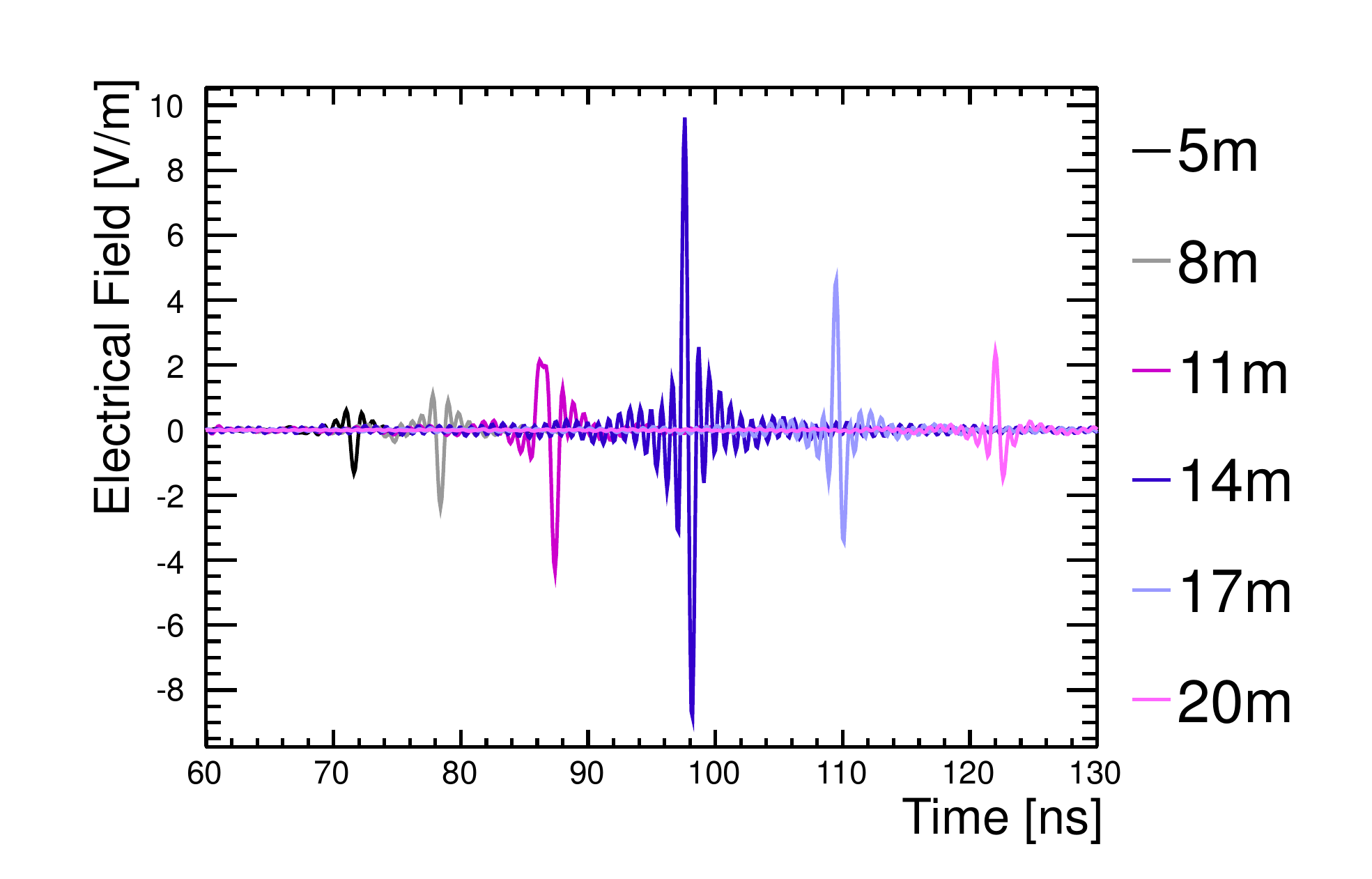} }\\
\vspace{0cm}
%
\centering{
%
\includegraphics[width=0.285\textwidth]{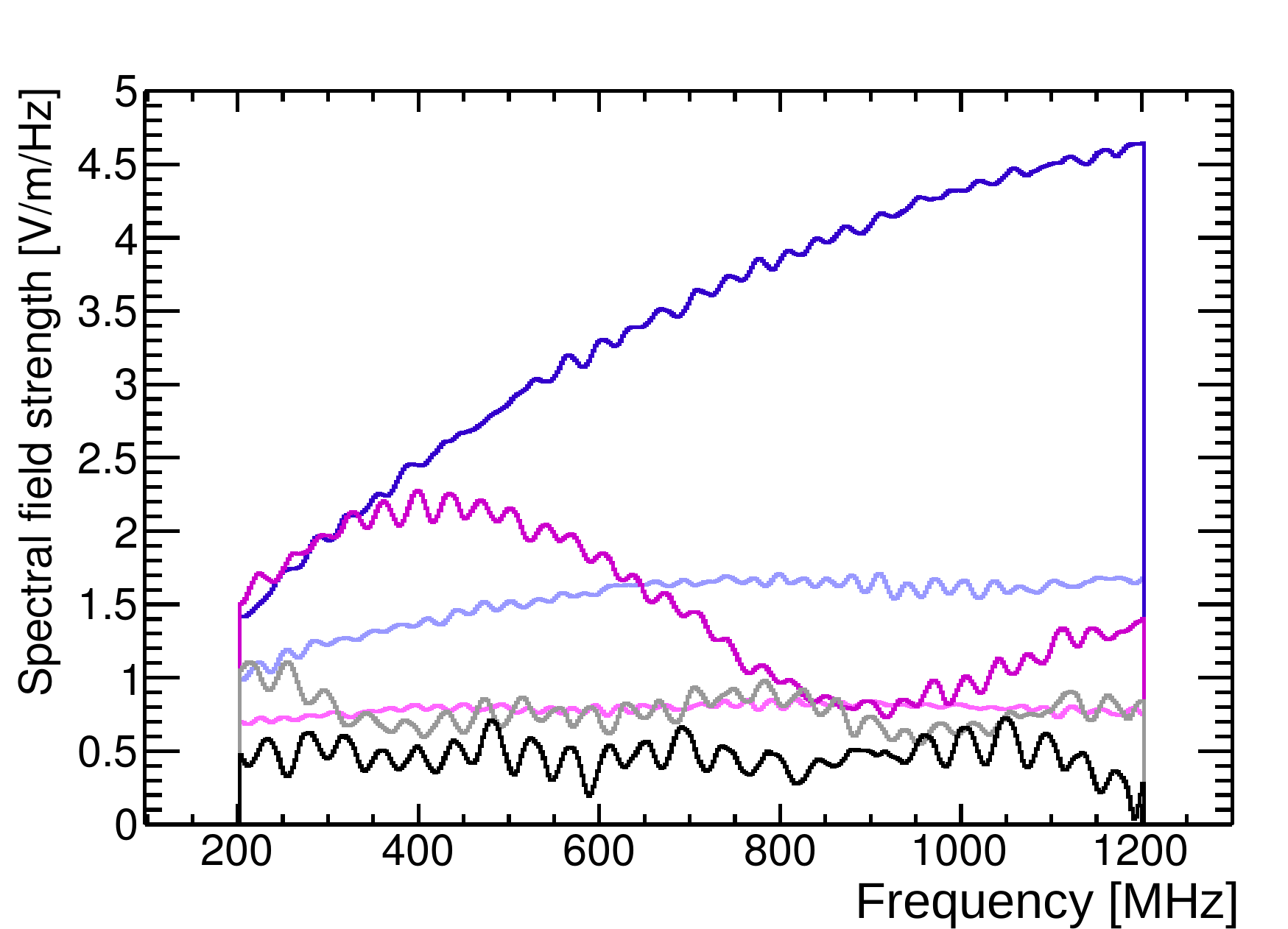}
%
\hspace{0.5cm}
%
\includegraphics[width=0.32\textwidth]{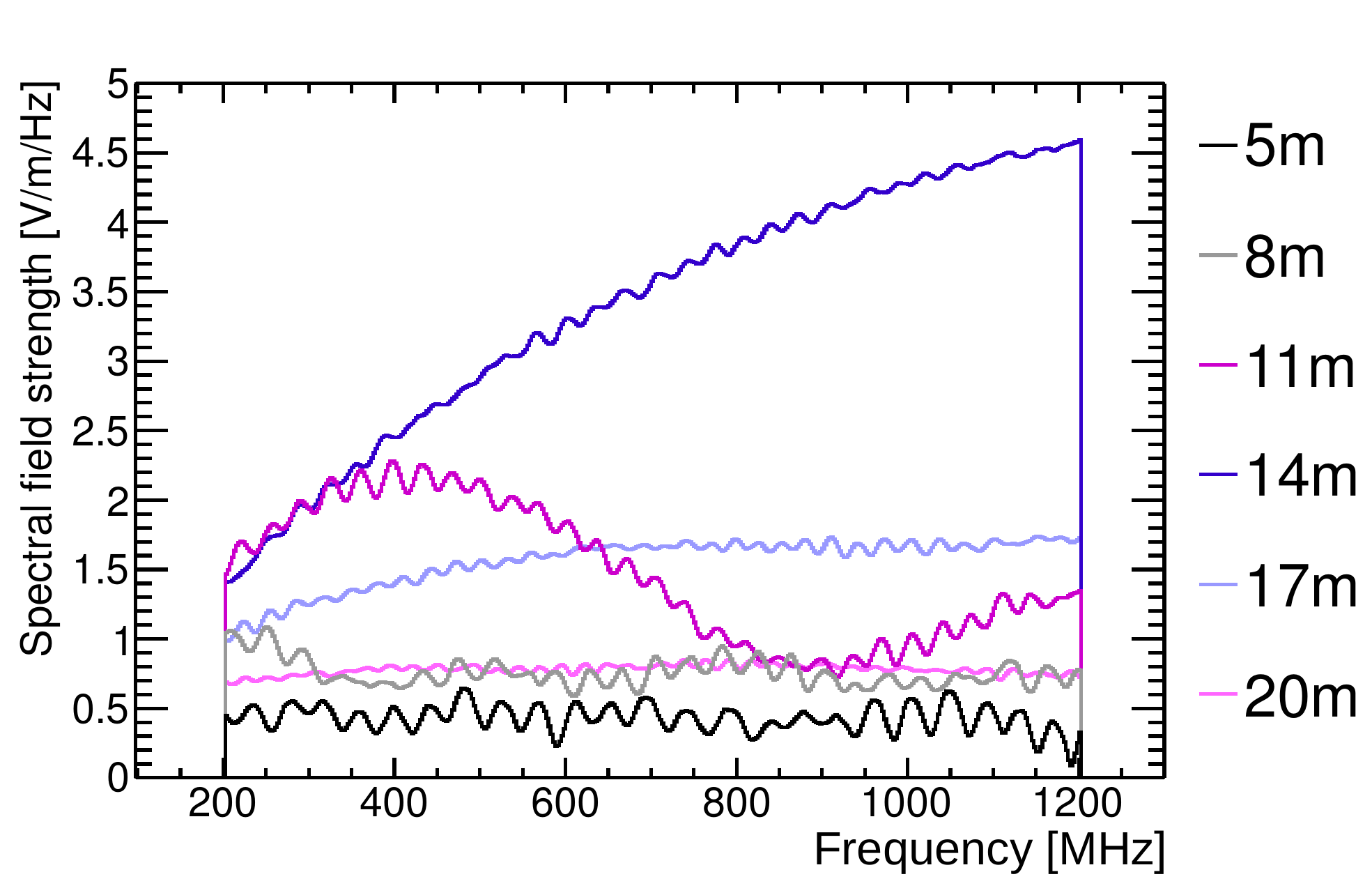}} 
\caption{Calculated electric field in the vertical component using the \enquote{end-point} (left) and \enquote{ZHS} (right) formalisms for different antenna positions $5-20\;\mbox{m}$ 
over the beam axis in a vertical slice along the center of the beam: top: in time domain; bottom: in frequency domain.}
\label{fig:Endpoint}
}
\end{figure}

\noindent
The simulation toolkit Geant4.9.6 is used for the simulation studies of this experiment. We take into account the details of the experimental setup, such as target geometry and material, as well as the beam particles and their energy. 
All relevant interactions of shower photons and electrons are included in the simulation. 
Using the feature of setting maximum subtrack size, no subtrack exceeds the length of $1\;\mbox{mm}$. 
The simulations include the calculation of the radio signals produced by the particle showers in the target based on the subtrack positions and times given by Geant4.  
Each subtrack contributes to the electric field or to the vector potential. 
Calculations using the \enquote{end-point} and \enquote{ZHS} formalisms run in parallel to provide a one-to-one comparison so that shower-to-shower fluctuations are not an issue.
Refraction at the target boundary is not yet taken into account in the propagation of the radio signals.
The simulations are done by injecting $500\;\mbox{electron primaries}$ with an energy of $4.35\;\mbox{GeV}$ each. The resulting electric field is then scaled up to a primary particle charge of $134\;\mbox{pC}$,
which is the average charge of the electron bunches measured during T-510 data taking.
The following simulations are done without a magnetic field present.
The radio signals are calculated for antennas positioned $9\;\mbox{m}$ from the end of the target and from $5-20\;\mbox{m}$ vertically above the beam axis.
A rectangle filter is applied for frequencies between $200\;\mathrm{-}\;1200\;\mbox{MHz}$.\\
\\
Figure \ref{fig:Endpoint} (left) shows the vertical component of the electric field in the time domain (top) and in the frequency domain (bottom) using the \enquote{end-point} formalism.
In both plots a strong signal at the Cherenkov angle at $14\;\mbox{m}$ elevation is visible, which can be explained by the Askaryan effect.
The position of the maximum amplitude of the signal agrees with expectation given by Cherenkov radiation for an refraction index of $n_{\text{HDPE}}=1.52$.

\begin{figure}[t]
\begin{minipage}[t]{0.29\textwidth}
\centering
\includegraphics[width=\textwidth]{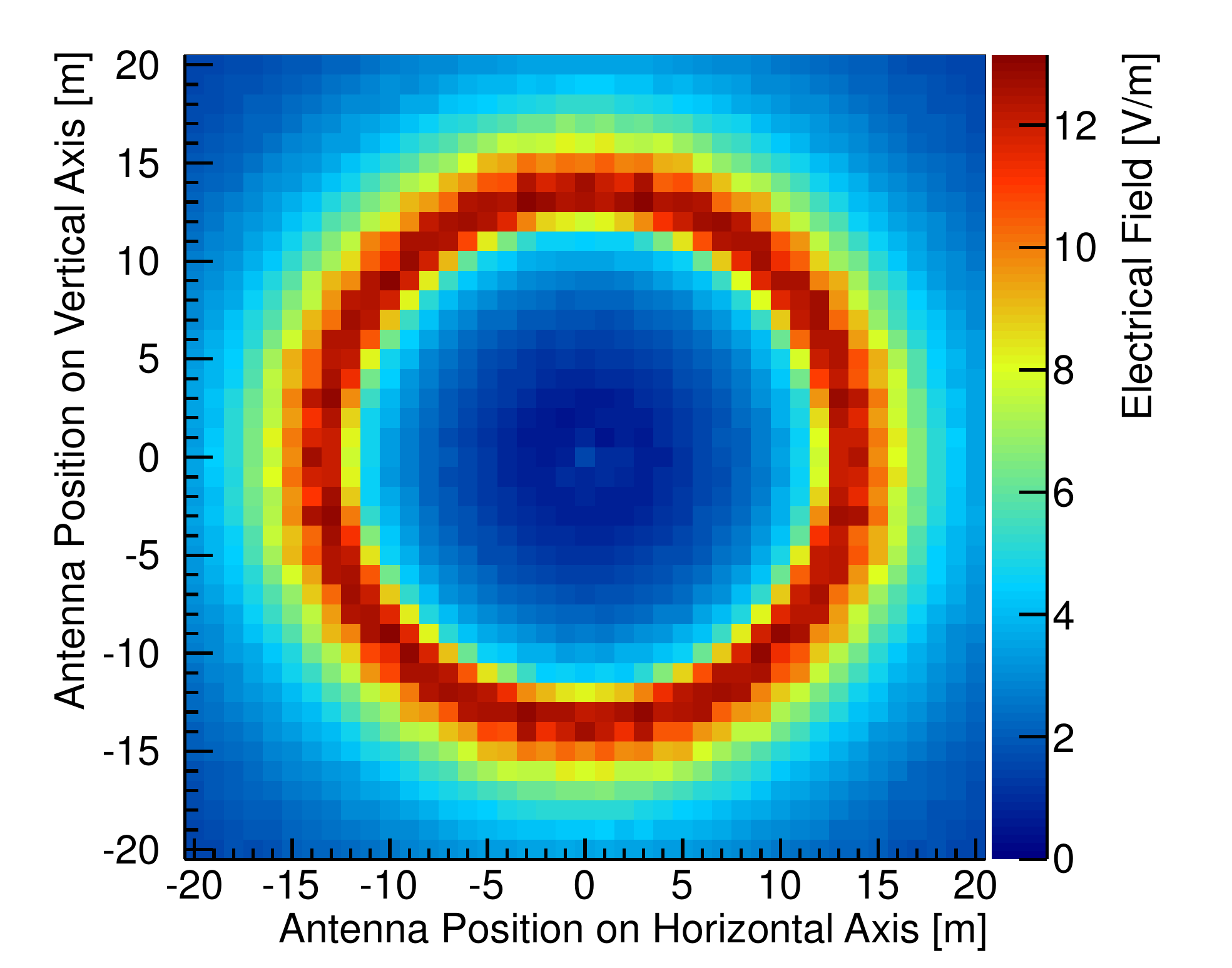}
        \caption{Caption for image}
        \label{fig:Endpoint_2D_total}
\end{minipage}
%
%
\begin{minipage}[t]{0.29\textwidth}
\centering
\includegraphics[width=\textwidth]{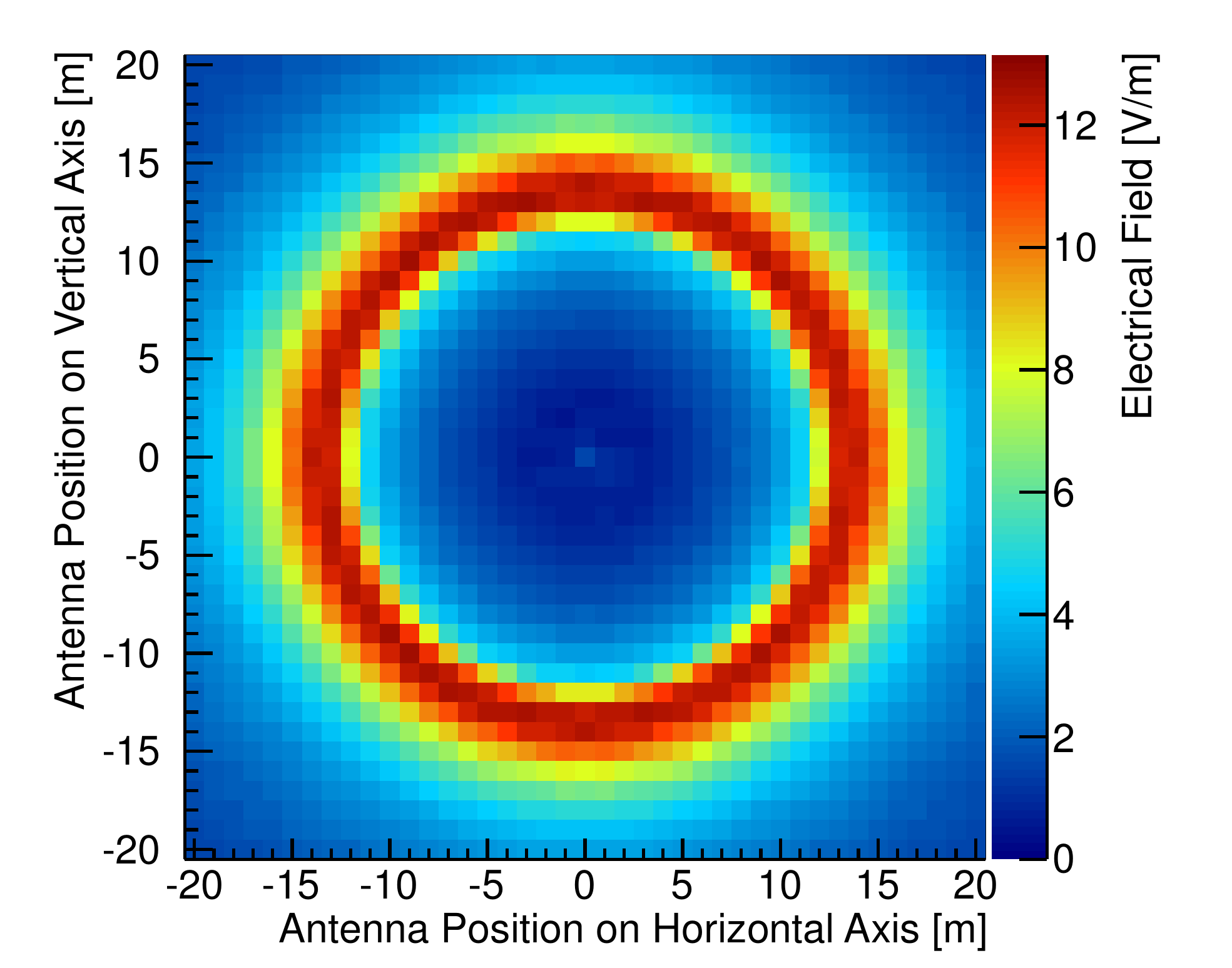}
        \caption{Caption for image}
        \label{fig:ZHS_2D_total}
\end{minipage}

\begin{minipage}[t]{0.29\textwidth}
\centering
\includegraphics[width=\textwidth]{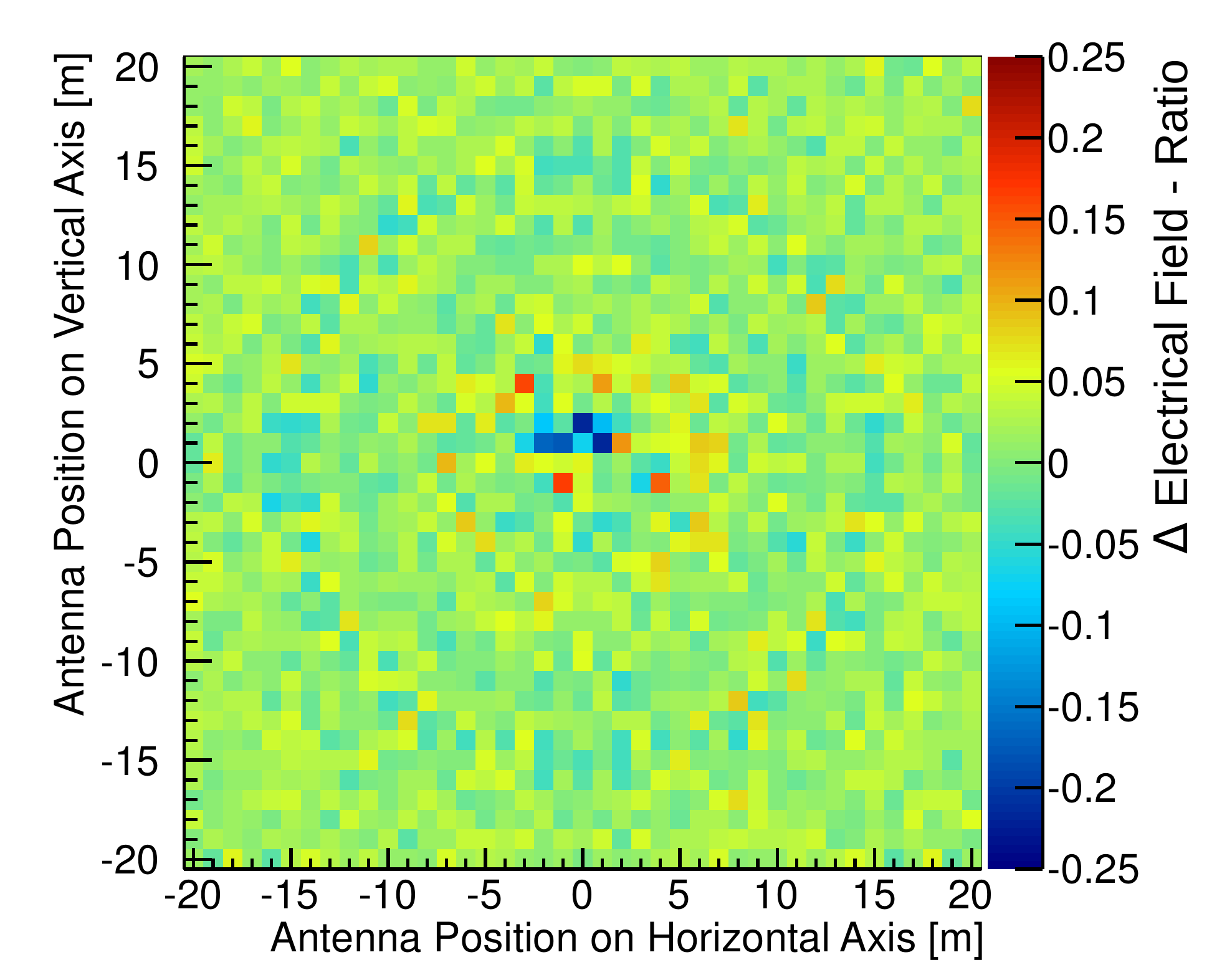}
        \caption{Difference}
        \label{fig:Diff_2D_total}
\end{minipage}
\caption{Peak amplitude of total electric field in time domain for a 2D antenna field:\\ left: using \enquote{end-point} formalism; center: using \enquote{ZHS} formalism; right: relative deviation of both formalisms.}
\label{fig:2D_total}
\end{figure}

\begin{figure}[b]
\begin{minipage}[b]{0.35\textwidth}
\centering
\includegraphics[width=\textwidth]{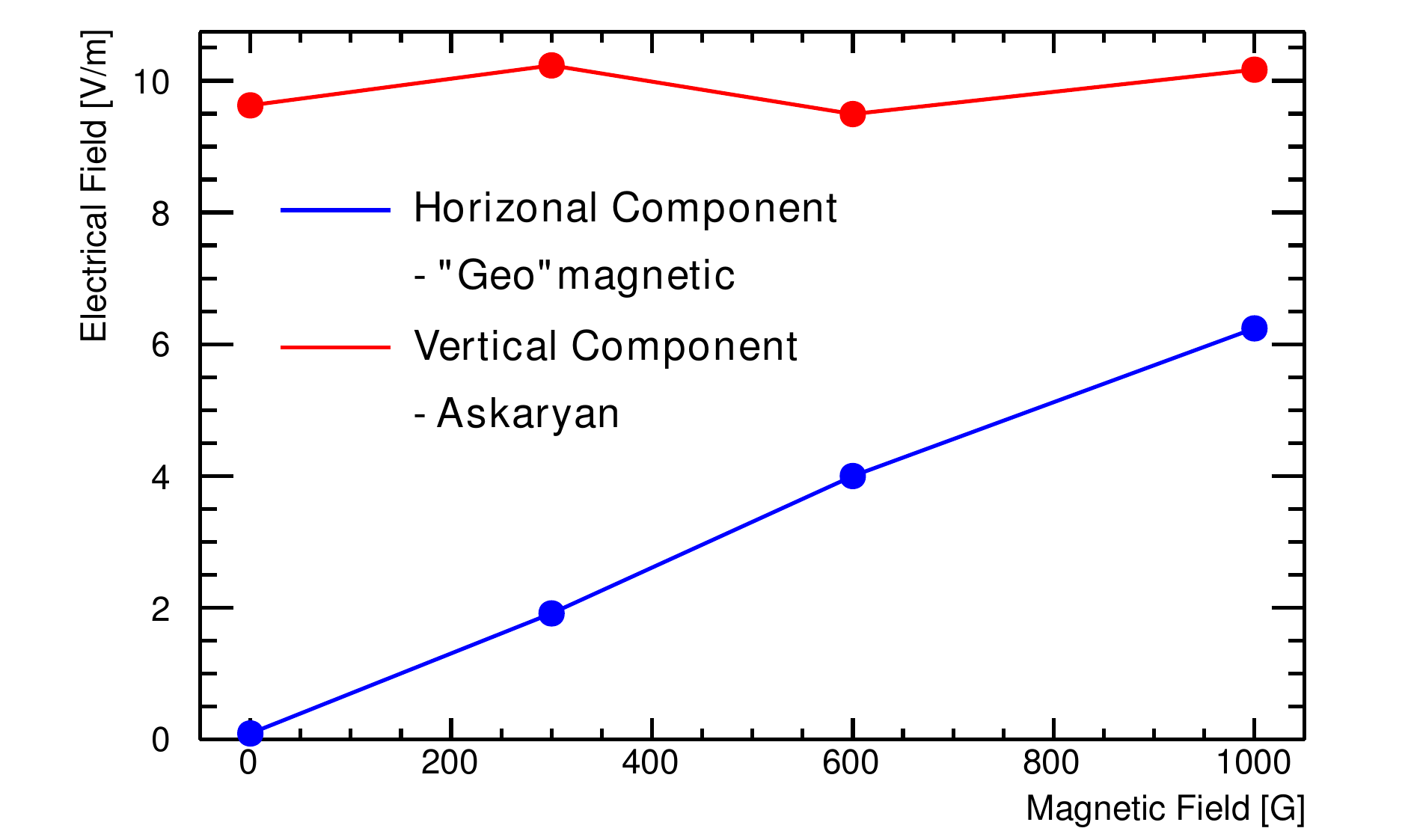}
        \caption{Signal dependency of the vertical and horizontal maximum electric field amplitudes on the magnetic field strength (\enquote{end-point} formalism).}
        \label{fig:xy}
\end{minipage}
\end{figure}

\begin{figure}[h]
\vbox{
\centering
\includegraphics[width=0.28\textwidth]{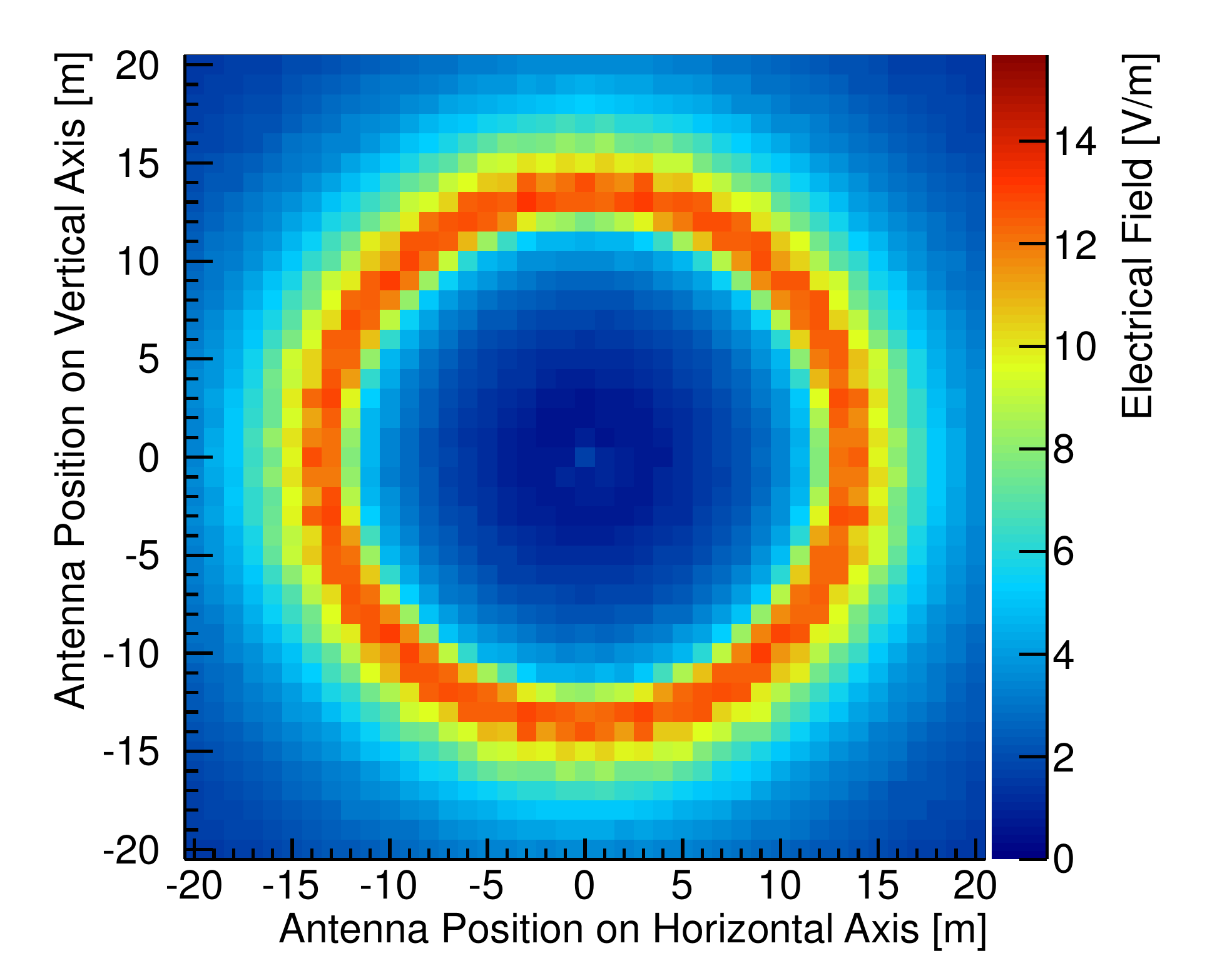}
%
%
\centering
\includegraphics[width=0.28\textwidth]{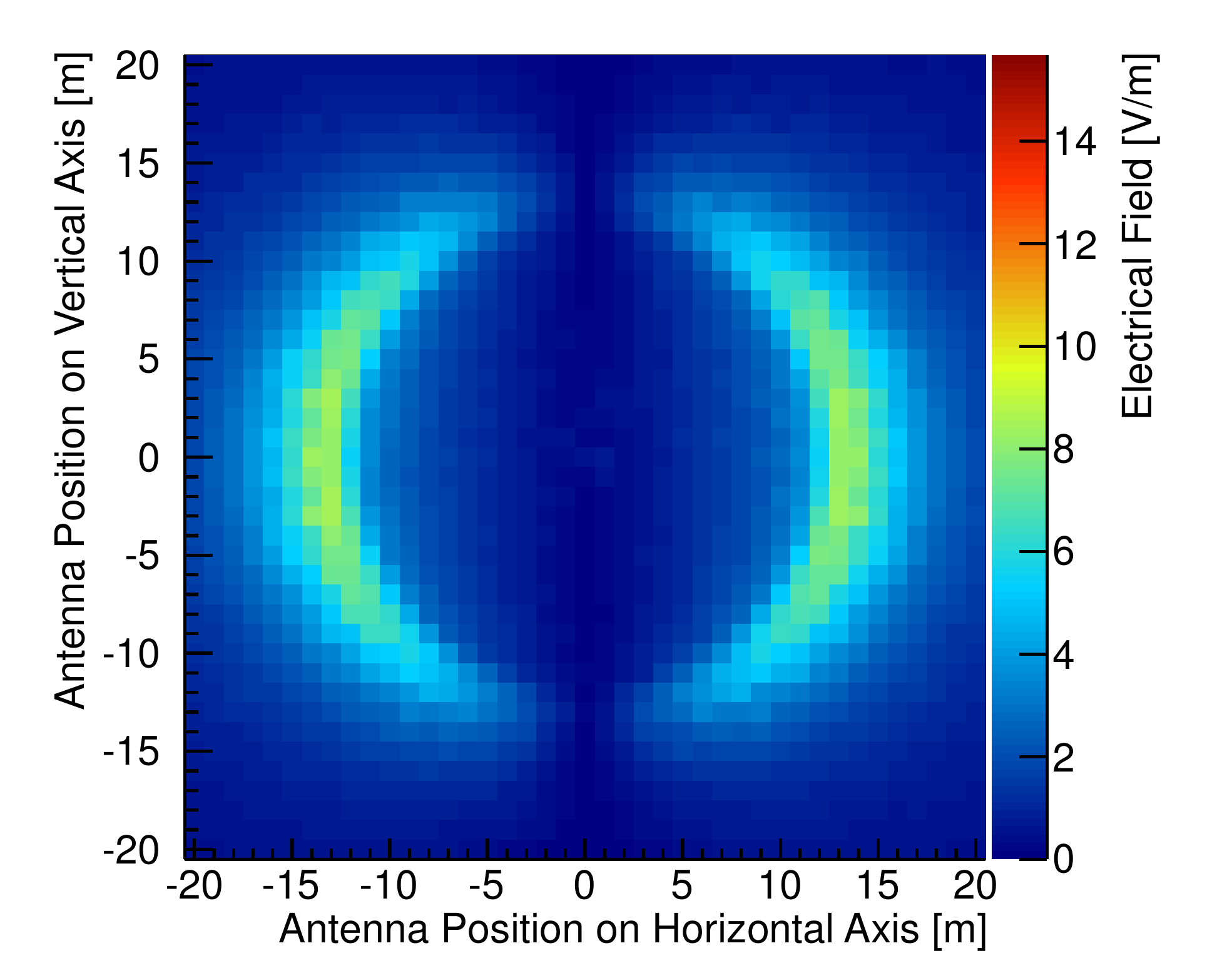}
%
%
\centering
\includegraphics[width=0.28\textwidth]{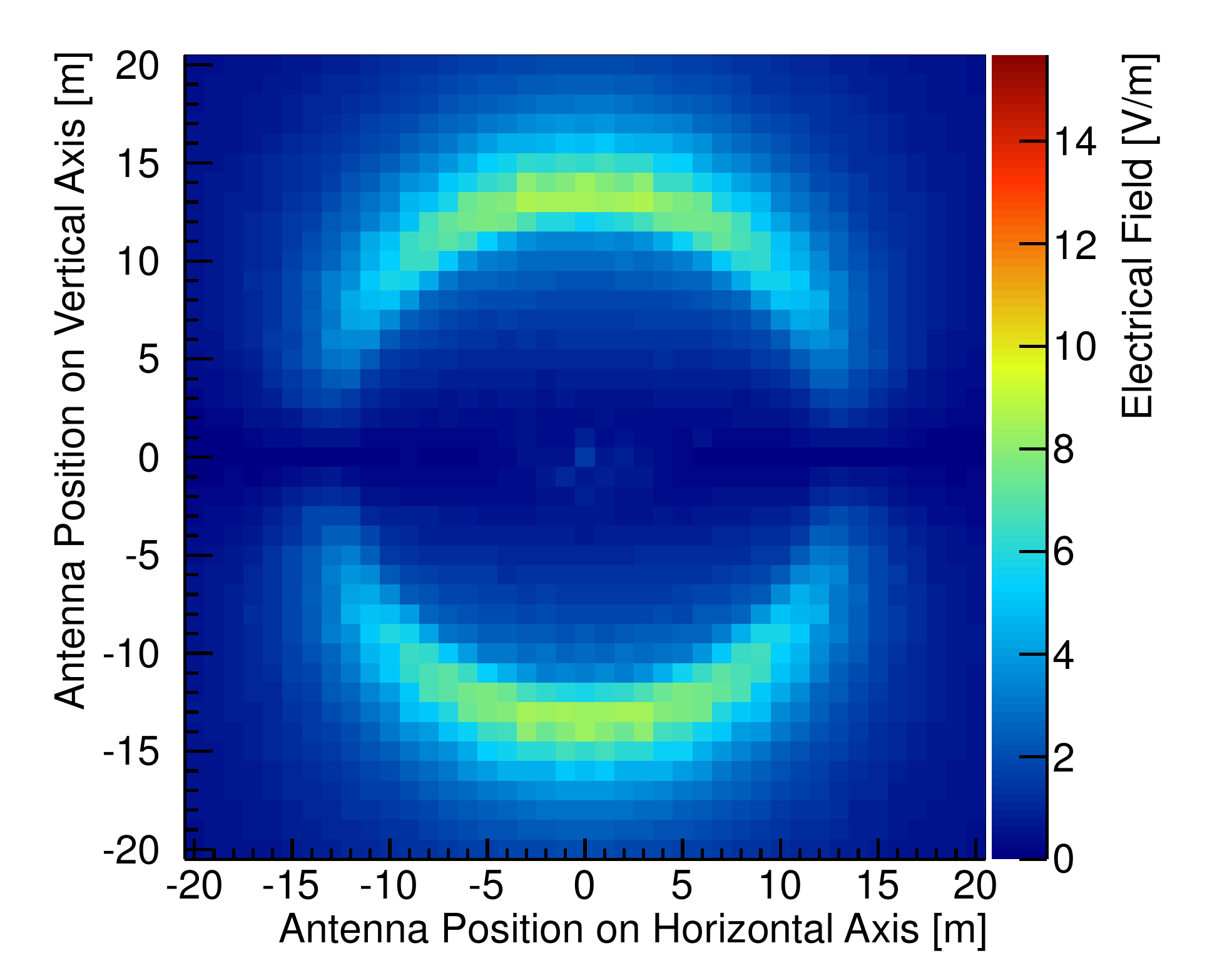}\\
\vspace{0.cm}
\centering
\includegraphics[width=0.28\textwidth]{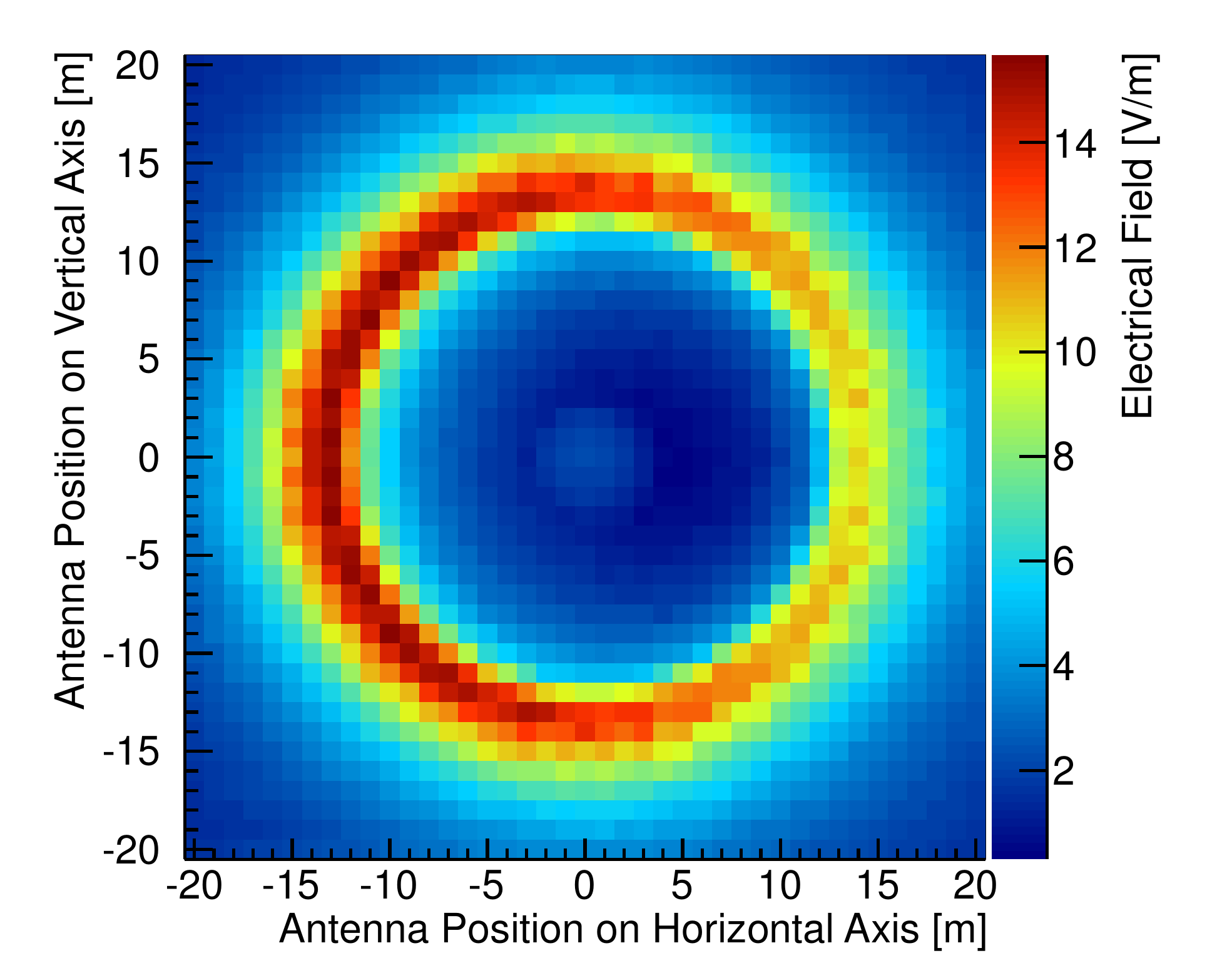}
%
%
\centering
\includegraphics[width=0.28\textwidth]{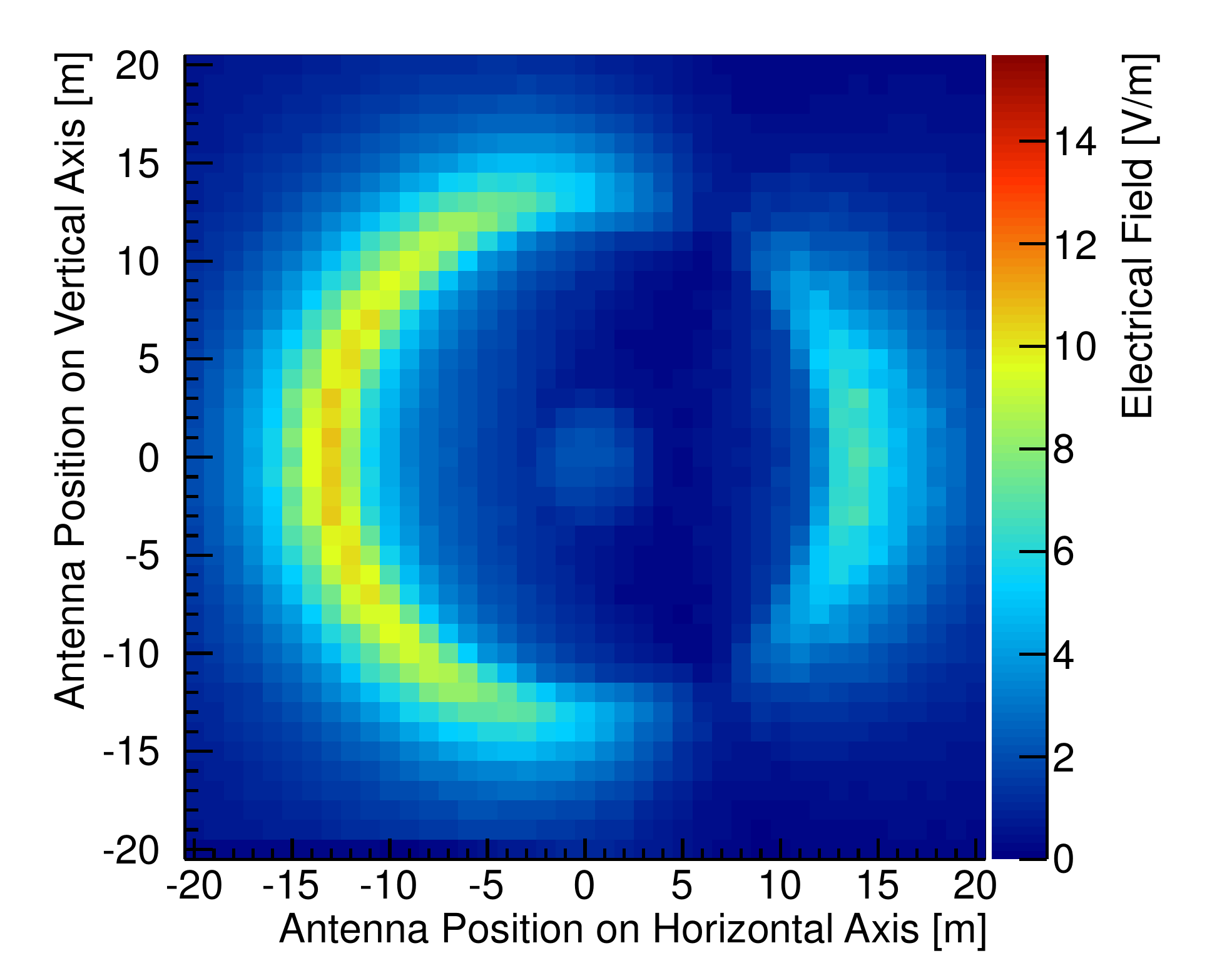}
%
%
\centering
\includegraphics[width=0.28\textwidth]{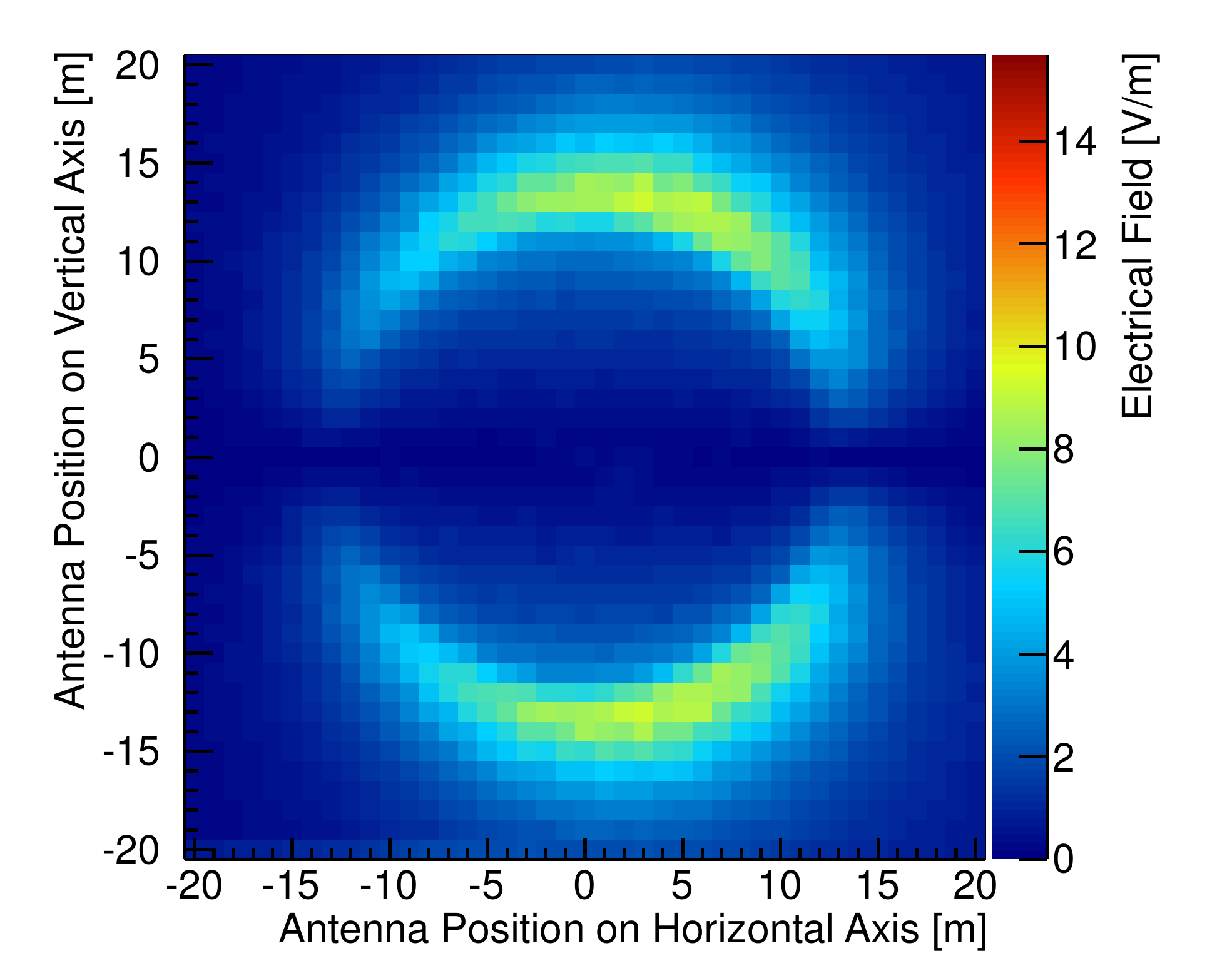}
\caption{Peak amplitude of electric field for a 2D antenna array using \enquote{end-point} formalism for a uniform magnetic field of a strength of $0\;\mbox{Gauss}$ (top) and $1000\;\mbox{Gauss}$ (bottom): left: total electric field; center: horizontal component right: vertical component.}

\label{fig:2D_total_B}
 }
\end{figure}

\noindent
For comparison, figure \ref{fig:Endpoint} (right) shows the vertical component of the electric field in the time (top) and frequency (bottom) domain  for the \enquote{ZHS} formalism. 
In this approach, the vector potential calculated by this formalism has been used to derive the electric field.
The amplitudes of the electric fields in the time domain, as well as the strengths of the electric field in the frequency domain are nearly identical for both formalisms.
The pulse shapes are very similar, and the \enquote{ZHS} formalism also has  a strong signal at the Cherenkov angle, which are in agreement with expectations.\\
\noindent
For a more detailed comparison, the electric field in the time domain has been calculated for an array of antenna locations, where the distance between the individual 
antennas is in $1\;\mbox{m}$ steps in positions horizontal and vertical to the beam axis. 
The distribution of the maximum amplitudes of the resulting total electric field for both formalisms is shown in figure \ref{fig:2D_total} for the \enquote{end-point} (left) 
and \enquote{ZHS} (center) formalisms. 
At first view, a strong Cherenkov ring is formed for both formalisms, and only small fluctuations are visible.
Figure \ref{fig:2D_total} (right) shows the relative deviation of the 2D distribution for the \enquote{end-point} formalism to the one for the \enquote{ZHS} formalism,
where a ring structure is visible in the distribution. However, the deviation in both directions are not larger than $5\;\%$.\\
\\
\noindent
\textbf{Influence of a magnetic field.} The SLAC T-510 experiment is capable of manipulating the intensity of the signal strength of the horizontal component of the electric field signal by changing a magnetic field along the vertical axis from the beam axis \cite{Katie}. 
The measured electric field strength in the horizontal component should rise linearly 
with rising magnetic field strength, due to the \enquote{geo}-magnetic effect, whereas the signal in the vertical component due to the Askaryan effect should be independent of the magnetic field.\\
To fully analyze the impact of the \enquote{geo}-magnetic effect, one has to study the influence of a magnetic field on the radio signal of particle showers in detail. Therfore, in Geant4 simulation, a uniform magnetic field is switched on which has
the strength up to $1000\;\mbox{Gauss}$ in the vertical direction perpendicular to the electron beam. Figure \ref{fig:xy} shows the results for the simulations in the presence of different uniform magnetic field strengths in the vertical direction 
perpendicular to the electron beam: $0\;\mbox{Gauss}$, $300\;\mbox{Gauss}$, $600\;\mbox{Gauss}$ and $1000\;\mbox{Gauss}$. Therfore, 
The maximum amplitudes in time domain of the electric field are sampled at antenna positions at the center of the Cherenkov cone and plotted versus the magnetic field strength. 
The expected behavior for both components as described previously is obtained to within the uncertainty due to statistical fluctuations.
\\
\noindent
The 2D distribution for the maximum amplitude of the total electric field under the influence of $1000\;\mbox{Gauss}$ magnetic field is shown in figure \ref{fig:2D_total_B} (bottom left). 
The \enquote{end-point} formalism is shown in figure \ref{fig:2D_total_B} (bottom left), and, for comparison,
the same distribution without a magnetic field from figure \ref{fig:2D_total} (top left) is shown again, with both plots scaled equally on the z-axis. 
Figure \ref{fig:2D_total_B} shows also the results for the horizontal (center) and vertical (right) components without (top) and with (bottom) the magnetic field present.
The Cherenkov ring structure is also present with the applied magnetic field, but it leads to an asymmetric rising of the signal amplitude on the ring structure to the left. 
This is observable especially in the horizontal component of the electric field (s. fig. \ref{fig:2D_total_B} center). 
The results for the \enquote{ZHS} formalism are not shown, as they are are nearly identical to those for the \enquote{end-point} formalism.\\
In a macroscopic view, we can describe this behavior via the superposition of the two main radio emission mechanisms of a particle shower: the \enquote{geo}-magnetic effect and the Askaryan effect.\\
\section{Conclusions}
\noindent
There are two established formalisms for the simulation of radio emission by extensive particle showers. 
To validate these formalisms, the SLAC T-510 experiment was performed to approximate air shower conditions in a controlled laboratory environment where the impact of 
the \enquote{geo}-magnetic effect could be examined.
First results of Geant4 simulations using both formalisms in parallel were shown and the results are found to be consistent, and the Cherenkov ring structure is clearly visible in the distribution of the electric field components.
In addition, a strong signal dependence on the magnetic field strength is shown using the \enquote{end-point} formalism.

Next, we will implement refraction at the boundary of the target and compare the modified simulation directly with the measurements of the SLAC T-510 experiment.\\
\\
\noindent
\textbf{Acknowledgments.} This experiment is supported in part by Department of Energy contract DE-AC02-76SF00515. 
This work got financial support by the Karlsruhe School of Elementary Particle and Astroparticle Physic (KSETA). 
The author would like to thank Carl Hudspeth and Janice Nelson for their support during data taking at SLAC.

\bibliographystyle{aipproc}
\bibliography{arena2012}

\end{document}